\documentclass[preprint,showpacs,preprintnumbers,amsmath,amssymb]{revtex4}
\usepackage{amsmath,amssymb,graphicx}
\usepackage {amssymb}
\newcommand{\nc}{\newcommand}
\nc{\ba}{\begin{eqnarray}}
\nc{\ea}{\end{eqnarray}}
\nc{\ga}{\gamma}
\def\ap{{\alpha^{\prime}}}

\input{epsf.sty}

\begin{document}


\title{Cosmological Perturbations on a Bouncing Brane }

\author{Robert Brandenberger}
\email{rhb@physics.mcgill.ca}
\author{Hassan Firouzjahi }
\email{firouz@physics.mcgill.ca}
\author{Omid Saremi}
\email{omid@physics.mcgill.ca}
\affiliation{Physics Department, McGill University, 3600 University Street,
Montreal, Canada, H3A 2T8 }

\begin{abstract}
The cosmological perturbations on a bouncing brane are studied. The brane is
moving inside a Klebanov-Strassler throat where the infra-red
region of the geometry is
smoothly cut off. For an observer confined to the world-volume of the brane,
this results in a non-singular bouncing mirage cosmology. We have calculated
the scalar perturbations corresponding to the normal displacements of the
brane. This is performed in the probe brane limit where the gravitational
back-reaction of the brane on the bulk throat is absent. Our model provides
a framework for studying the transfer of fluctuations from a contracting
to an expanding phase. We find that the spectral index of the dominant mode
of the metric fluctuation is un-changed, unlike what is obtained by
gluing contracting to expanding Einstein universes with the help of
the usual matching conditions. Assuming that the fluctuations
start off in a vacuum state on sub-Hubble scales during the contracting
phase, it is shown that the resulting spectral index $n_s$ on super-Hubble
scales in the expanding phase has a large blue tilt. When the brane is
moving slowly inside the throat and its kinetic energy is negligible
compared to its rest mass, one finds $n_s=3$. For a fast-rolling brane
with a large kinetic energy, the spectral index is $n_s = 2.3$.
This may put severe constraints on models of mirage cosmology.

\vspace{0.3cm}

\end{abstract}

\maketitle

\section{Introduction}

Recent cosmological observations provide an increasingly clear picture
of the current structure of the universe on large scales
\cite{Spergel:2006hy}. The universe is spatially flat to high
accuracy, the background cosmology is homogeneous and isotropic, and
there is a superimposed spectrum of almost scale-invariant,
nearly Gaussian, and nearly adiabatic cosmological fluctuations.
A viable theory of early universe cosmology must explain these observations.
Inflationary cosmology \cite{guth} by far is the best model currently
available which can explain these data. Furthermore, it successfully solves
the flatness, homogeneity and horizon problems of Standard Big Bang (SBB)
cosmology. On the other hand, current realizations of inflation are plagued
by various conceptual problems \cite{RHBrev1}. Two of the problems which
motivate this study are the presence of an initial singularity \cite{Borde}
which signals an incompleteness of the background cosmology, and the
``Trans-Planckian'' problem, namely the fact that the fluctuations emerge
from sub-Planckian wavelengths and hence from a zone of ignorance about
the fundamental physics \cite{RHBrev1,Martin}. One should thus be open-minded
towards alternative scenarios which may, in principle, be able to explain
the data while avoiding the shortcomings associated with inflationary
cosmology.

Bouncing cosmologies can solve the horizon problem of the SBB cosmology.
In bouncing cosmologies, each period of expansion is preceded by a period
of contraction during which the co-moving Hubble radius is decreasing while the
horizon continues to increase. Hence, in a bouncing universe the horizon
can easily be made to be larger than the past light cone at the time of
last scattering, the region over which the universe is seen to be homogeneous
and isotropic. The entropy or size problem of SBB cosmology also disappears
in the context of a bouncing cosmology: if the universe begins large in a
phase of contraction, there is no reason why the initial entropy should be
small. The flatness problem of SBB cosmology, however, persists
(see e.g. \cite{pyro}).

A key question which determines the viability of any cosmological model
with a bounce concerns the spectrum of cosmological perturbations. It
is an extremely non-trivial challenge to find a scenario which is consistent
with the current data. There are two issues involved. First, one must
find a mechanism for producing an almost scale-invariant spectrum, and
secondly, one must be able to evolve the spectrum through the bounce
region.

Recently, there have been several attempts to construct bouncing cosmologies,
in particular the Pre-Big-Bang \cite{PBB} and the Ekpyrotic \cite{EKP}
scenarios. Both, however, involve a singularity at the bounce point
which prevents a consistent computation of the spectrum of fluctuations.
(In a new version of the Ekpyrotic scenario, the singularity
can be smoothed out by invoking ghost condensation \cite{NewEkp}.)
There have been attempts to construct non-singular bouncing cosmologies,
making use of the interplay of curvature and matter with wrong sign
kinetic terms \cite{Peter}, quintom matter \cite{quintom}, k-essence
\cite{Abramo} and higher derivative gravity actions \cite{Tsujikawa,Biswas}.
An elegant way of obtaining a non-singular bouncing cosmology is by
means of ``mirage cosmology'' \cite{mirage}.

In this paper, we study the evolution of fluctuations through a
non-singular bounce in the context of mirage cosmology. We focus
mainly on two questions. First, does the spectral index of the
cosmological fluctuations change when passing through this bounce?
Secondly, does a scale-invariant spectrum emerge if we set up the
fluctuations in a vacuum state early in the period of contraction.
The answer to the second question is ``no''. Concerning the first
question, we find that the spectrum goes through the bounce without
change in the spectral index. This result is very interesting. In
work on Pre-Big-Bang and Ekpyrotic cosmology where the contracting
phase and the expanding phases were each described by dilaton or
Einstein gravity and connected at the singularity by matching
conditions \cite{Hwang,Deruelle} analogous to the Israel matching
conditions \cite{Israel}, it was found \cite{PBBflucts,Ekpflucts} that the
spectral index changes: the dominant mode of the metric fluctuations
in the contracting phase couples only to the decaying mode in the
expanding phase. However, the applicability of these matching
conditions has been challenged \cite{Durrer} and concrete
studies \cite{Peter2,ABBM,Abramo} have shown that it is possible
that the spectral index does not change if the bounce is smooth
(although other studies give differing results \cite{other,Tsujikawa}).
Mirage cosmology provides a simple setting to study
the issue of the transfer of cosmological perturbations through
a bounce in a clean way.

\section{The bouncing background}

In ``mirage cosmology''  \cite{mirage} our universe is a probe
D3-brane moving in extra dimensions.  Due to the motion of the brane
in the extra dimensions, the induced metric on the brane becomes
time-dependent. For an observer confined to the brane this results in a
cosmological expansion or contraction. In the probe brane limit, we
neglect the back-reaction of the moving brane on the background.
This is a good approximation when a single brane is moving in the background
of a large stack of branes or in the background created by a large number
of flux quanta.

The bouncing brane in a throat has been studied in
\cite{Kachru:2002kx}, \cite{Germani:2006pf}, \cite{Easson:2007fz}
and \cite{Germani:2007ub}. The throat was taken to be the warped
deformed conifold of the Klebanov-Strassler (KS) solution
\cite{Klebanov:2000hb}, where the infra-read (IR) region of the
geometry is smoothly cut off.  The ultra-violet (UV) region of the
throat is smoothly glued to the bulk of the Calabi-Yau (CY)
manifold. The brane sets off from the UV region, moving towards the
IR region. For an observer on the brane this corresponds to a
contracting phase. At the tip of the throat, where the warp factor
is stationary, the induced scale factor becomes stationary. This
results in a bounce point. Finally, the brane bounces back to the UV
region which corresponds to an expanding period. In order for the
scenario to work, then one has to find a way to consistently connect
this period to a late time Big Bang cosmology.

One novel feature of this bouncing solution is that it is singularity-free.
This is a manifestation of the fact that the IR region is smoothly cut off.
Furthermore, in \cite{Germani:2006pf} and \cite{Easson:2007fz} it is shown
that a cyclic cosmology can be achieved by turning on internal angular
momenta. With angular momenta turned on, the brane motion is confined
between IR and UV region of the throat which results in a cyclic solution.

The metric of the background is given by
\ba
\label{metric1}
ds^{2} \, =     G_{MN}  \, d\, X^{M} \, d\, X^{N}    \, =  h(r)^{-1/2} \eta_{\mu \nu} d\, x^{\mu}  d\, x^{\nu}
+ g(r) d\, r^{2} + ds_{5}^{2} \, .
\ea
Here, $G_{MN}$ is the background metric, $h(r)$ is the warp factor, $r$ is the radial direction of the
throat and $ds_{5}^{2}$ is the metric along the internal angular
directions. In what follows, we do not consider the angular motion
of the brane and neglect the last term in Eq. (\ref{metric1}).

The space-time indices are represented by capital Latin letters $M, N,...$
while the indices used by the observer confined to the brane are
represented by  $a,b,...$. The brane is equipped with coordinate ${\xi^{a}}$.
In general, when the brane is moving non-uniformly, one cannot use the
static gauge and $\{ \xi^{a} \}\neq \{x^{\mu}\}$. However, the
fluctuations along the spatial coordinates of the branes, $\xi^{i}$, are
not physical and we can choose $x^{i}=\xi^{i}$ for $i=1,2,3$. The time
coordinate on the brane is defined by $\xi^{0}=t$. In the homogeneous limit
when the brane is moving uniformly and no perturbations are present on the
brane world-volume, we can set $X^{0}=t$.

The action of the probe D3-brane is given by
\ba
\label{action1}
S \, = \, T_{3}\, \int d^{4} \xi \sqrt{-|\ga_{ab}|} +
\mu_{3}\, \int_{{\cal M}} d^{4} \xi \, P(C_{(4)}) \, .
\ea
Here, the first term is the usual Dirac-Born-Infeld (DBI) term and the
second one is the Chern-Simons (CS) term. Also $T_{3}$ and $\mu_{3}$ are the
brane tension and charge, respectively, $\ga_{ab}$ is the induced metric
on the brane, $C_{(4)}$ is the background Ramond-Ramond
four-form field and $P(C_{(4)})$ represents its pull-back onto the world
volume ${\cal M}$ of the D3-brane. In the following, we consider the motion
of a BPS brane and set $\mu_3 = T_3$.

The position of the brane in the extra dimension is parameterized by
\ba
\label{XM}
X^{M} \, = \, X^{M}(\xi^{a}) \, .
\ea

For a background space-time metric $G_{MN}$ given in Eq.
(\ref{metric1}), we have
\ba
\label{gamma}
\ga_{ab} \, = \, G_{MN}\,  \frac{\partial{X^{M}}}{\partial{\xi^{a}}}
\frac{\partial{X^{N}}}{\partial{\xi^{b}}}
\quad , \quad
P\,( C_{(4)})_{t123}=  \frac{\partial X^{0}}{\partial t} {C_{(4)}}_{0123} \, .
\ea

For the KS background \cite{Klebanov:2000hb}, the radial part of the
metric and the form of the warp factor is given by (for a detailed
review of this background see \cite{Herzog:2001xk})
\ba \label{KS} g\, = \, \frac{2^{1/3}}{6} (g_{s} M \ap)\,
I(r)^{1/2}\, K(r)^{-2} \quad ,\quad h(r) \, = \, (g_{s} M \ap)^{2}
2^{2/3} \epsilon^{-8/3} I(r)\, , \ea
where
\ba
\label{IK}
I(r) \, = \, \int_{r}^{\infty} dx \frac{x \coth\, x -1}{\sinh^{2} \, x}\,
(\sinh \, 2x - 2x )^{1/3}
\quad , \quad
K(r) \, = \, \frac{ (\sinh \, 2r - 2r )^{1/3} } {2^{1/3} \sinh r} \, .
\ea
In these expressions, $g_{s}$ is the string coupling, $M$ is the quantum
number of the RR-fluxes turned on inside the $S^{3}$
at the bottom of the throat, and $\epsilon^{2/3}$ measures the size of
this $S^{3}$ in the units of $\ap$.
The warp factor $h(r)$ has a maximum at the bottom of throat $r=0$ and
falls-off exponentially at large $r$. The function $g(r)$ has a minimum at
$r=0$ and increases like $r^{1/2}$ for large $r$.
The fact that $h(r)$ and $g(r)$ are non-singular and their first derivatives
vanish at $r=0$ is crucial to obtain the bounce solution. In this background,
the RR four-form field is given by
\ba
\label{C4}
C_{(4)_{0123}} \, = \, g_{s}^{-1}\, h(r)^{-1}\, .
\ea

At the homogeneous level the brane moves along the radial directions without
any perturbation, and, using
Eqs. (\ref{gamma}) and (\ref{C4}), we obtain
\ba
\label{Action}
S \, = \, \int\, d^{4}\xi \, {\cal L}=  T_{3}\, \int\, d^{4} \xi\,\, h^{-1}\,
\left( - \sqrt{ 1- h^{{1/2} }  g\, {\dot r}^2 } +1 \right) \, ,
\ea
where the relation  $X^{0}=t$ was used. Here and in what follows, an over-dot
corresponds to a derivative with respect to $t$, the time measured by the
observer on the brane.

Integrating the radial Euler-Lagrange equation which follows from this
action, one can construct a constant of integration
\ba
E \, = \, \dot r \, \partial {\cal L}/ \partial \dot r - {\cal L} \, .
\ea
This yields the following first order equation of motion
for the position of brane $r= R(t)$
\ba \label{eom0} \dot R ^{2} + V_{eff}(R) \, = \, 0 \quad , \quad
V_{eff}(R) \, = \, -\frac{E\, h^{1/2} \, (2+ E\, h)}{g \left(1+E\,
h\right)^{2}} \, . \ea

It is easy to see that $E>0$ and $V_{eff}$ is always negative. Thus,
the brane can move in the entire region $0< r<\infty$. On the other
hand, to get a finite value for Newton constant (finite $M_{P}$),
the KS throat should be glued smoothly to the bulk of the CY
manifold at $r=r_{c}$. This implies that the brane would be moving
in the region $0 < r < r_{c}$. We restrict our analysis to the
throat region of the KS construction.

An observer confined to the D3-brane feels cosmological expansion or
contractions, depending on the direction of the brane's motion. For this
observer, the four-dimensional metric is given by
\ba
\label{4D}
ds_{4}^{2}  =  -d\tau^{2} + a(\tau)^{2}  \,    d{\bf x}^2  \, ,
\ea
where ${\bf x}$ denote the three spatial coordinates and the proper time is defined by
\ba
\label{tau}
d\, \tau^{2} \, = \, -\ga_{00}\,  d t^{2} = h^{-1/2} \,
(1- g\, h^{1/2} \dot R^{2} ) d t^{2}\, ,
\ea
and the scale factor is given by $ a(\tau) \, = \, h\left(R(t) \right )^{-1/4}$.

The corresponding Friedmann equation is
\ba
\label{Fried}
H^{2} \, = \, \frac{1}{16}
\frac{h'^{2}}{g\, h^{2}} \, ( E^{2} h^{2} + 2 E h) \, ,
\ea
where $H = \frac{1}{a(\tau)} \frac{d \, a(\tau)}{  d\, \tau}$
is the Hubble constant and $'$ represents the derivative with respect to $r$.
\begin{figure}[t] 
\vspace{-2cm}
   \centering
   \hspace{-1.8cm}
  \includegraphics[width=4.1in]{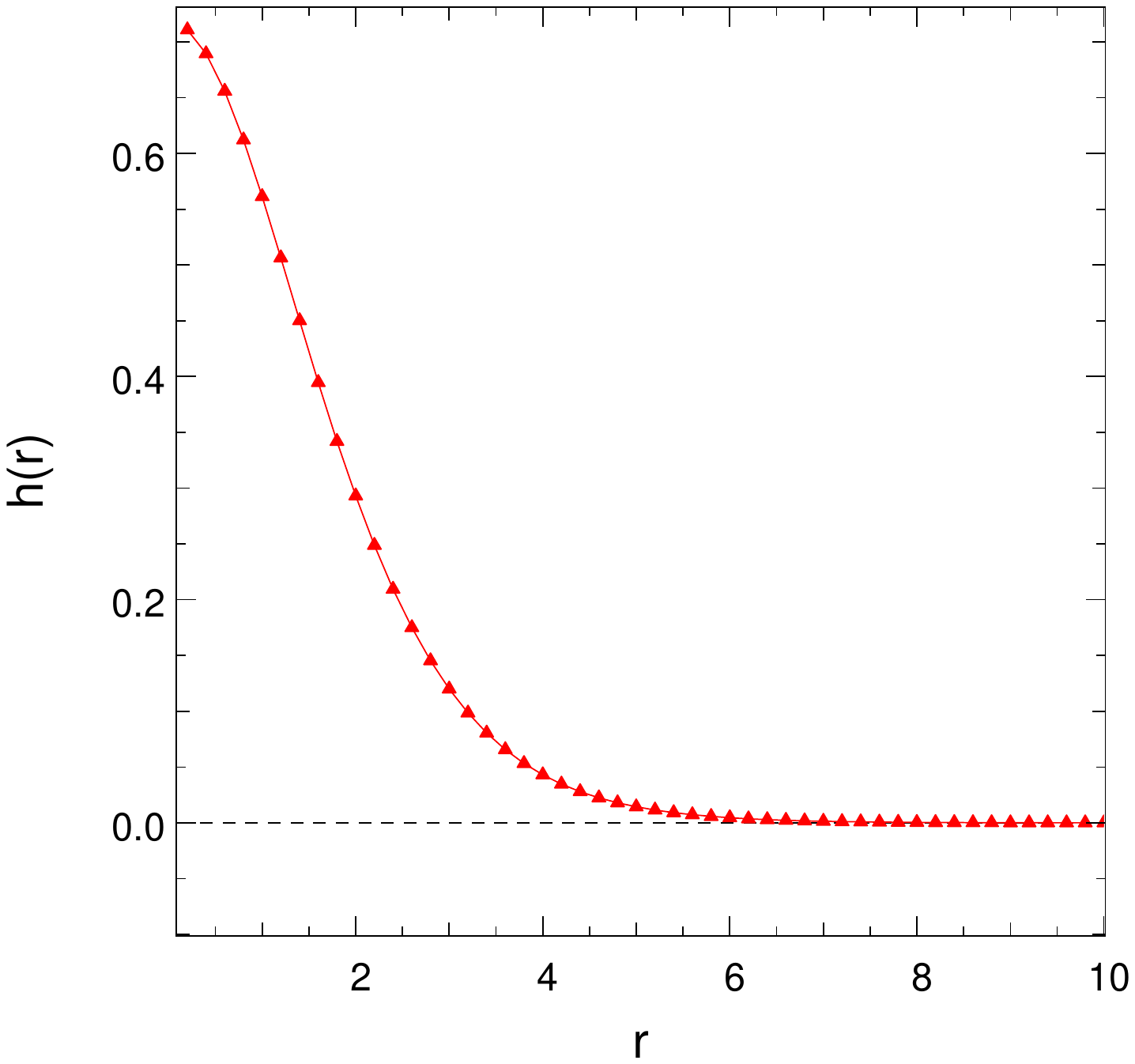} \hspace{-3.5cm}
    \includegraphics[width=4.1in]{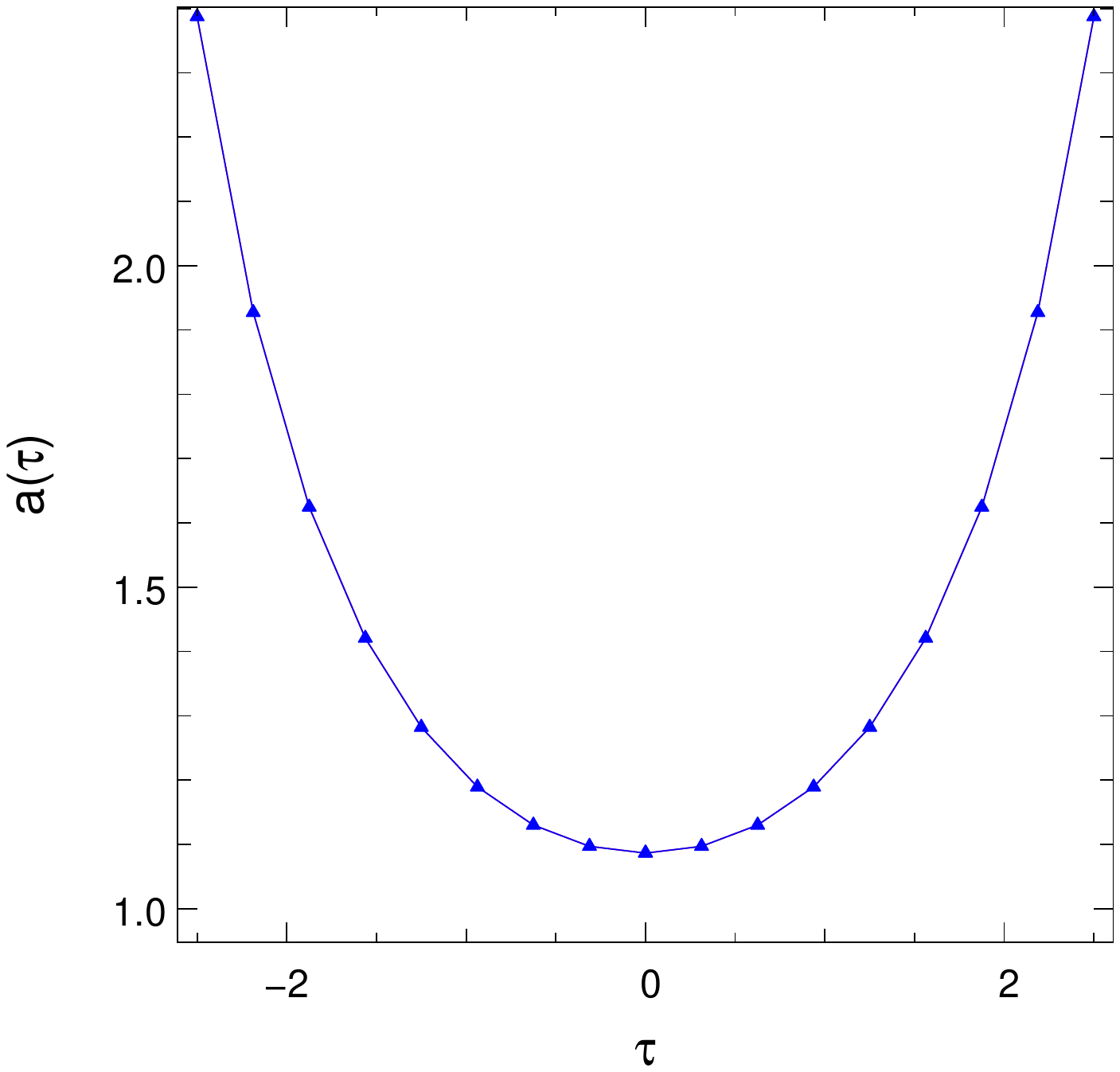} \hspace{-1cm}
    \vspace{-4.5cm}
\caption{The warp factor $h(r)$ and the scale
factor $a(\tau)$ are presented here. The scale factor is given by
$a(\tau)= h(R(t))^{-1/4}$. The contracting phase corresponds to the
case when the brane moves from the UV region (large $r$) towards the
IR region (small $r$). The bounce is at $r=0, \tau=0$, where $H=0$.
The expanding phase is the mirror image of the contracting phase. }
\vspace{0.8cm}
\label{bouncefig}
\end{figure}


{\bf Fig. \ref{bouncefig}} shows how the scale factor behaves as a function of
$\tau$ and $h$ as a function of $r$ (in $h(r)$ the prefactor has
been set to 1).
When the brane moves from the UV region to the IR region of the
geometry, the observer
sees a cosmic contraction. At $r=0$, $h'=0$, so $H=0$. This
corresponds to a bounce point \cite{Kachru:2002kx}. Afterwards,
the brane bounces back towards the UV region and the observer sees a
cosmic expansion. The expansion rate $H$ in the expanding phase is
the mirror image of the expansion rate in the contracting phase.
Using the asymptotic forms of the functions $I(r)$ and $K(r)$ given in
{\bf Appendix B}, one can see that immediately after the bounce, $H$ enters a
period of acceleration, followed by a period of deceleration.

To perform the mode analysis in the following sections, it is very helpful
to go to conformal time $\eta$, where$
a ^{2} d \eta^{2} \, = \, d \tau^{2}$. Using Eq. (\ref{eom0}), we obtain
\ba
\label{reta}
( \frac{d\, R}{d\, \eta} )^{2} \, = \,
\frac{ h^{1/2}}{ g} \, E\, (2+ E\,h) \, .
\ea

\section{Fluctuations of the Brane position}

In this section we study the fluctuation of the brane's position
while it is moving inside the throat.
We are particularly interested in the evolution of these perturbations
measured by the observer confined to the brane. Our study is a
generalization of what was previously studied
in \cite{Boehm:2002kf}.

\begin{figure}[t]
   \centering
   \hspace{-2.5cm}
   \includegraphics[width=3in]{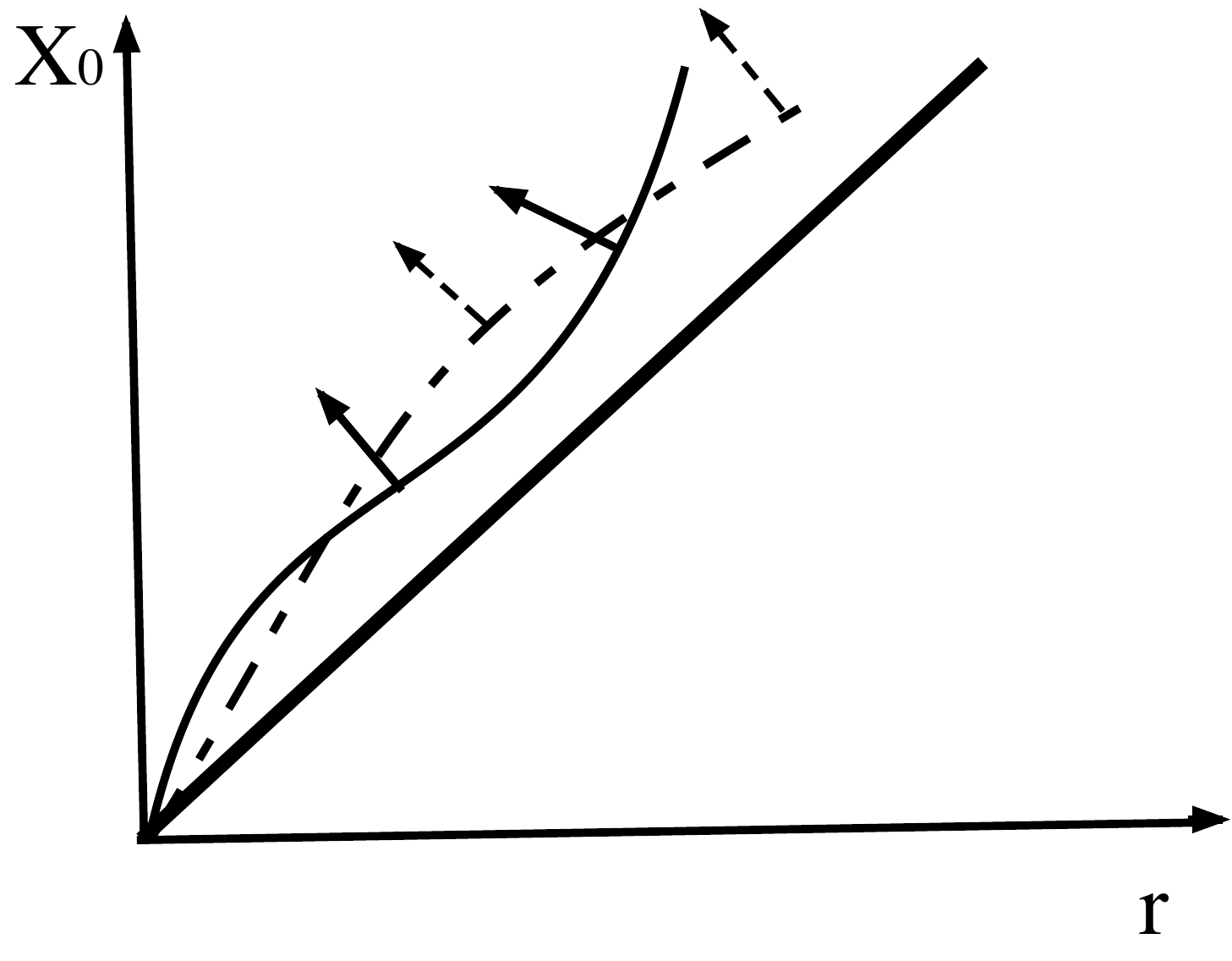} \hspace{-2.5cm}
\caption{In this plot the light cone for a fixed point on the brane is plotted. The dashed curved line represents the light-cone of the point when the brane is moving homogeneously governed by 
Eq. (\ref{eom0}). The unit normal vectors are in $X^{0}-r$ plane and have components in both $r$ and $X^{0}$ directions as is evident from Eq. (\ref{n}).
The curved line represents the light-cone when the normal perturbations Eq. (\ref{normal}) are present. Now $n^{M}$ also has components in the tangential directions $X^{i}$.}
\vspace{0.8cm}
\label{normalfig}
\end{figure}

In general, the displacements of the brane can be decomposed into
longitudinal and normal components. However, the longitudinal
perturbations (which are along the tangential directions of the brane)
are not physical and can be gauged away. Suppose $n^{\mu}$ is the
space-like unit normal vector to the brane. Then the normal
displacements of the brane are given by
\ba
\label{normal}
X^{M} \, = \, \bar {X }^{M} + \Phi\, n^{M}  \, .
\ea
Here, $\Phi$ is the scalar field corresponding to the normal displacement
and $\bar {X }^{M} $ represents the brane position at the homogeneous
background, i.e. the zeroth order brane position.
It is important to note that just considering perturbation along the
$r$ direction is not consistent, at least when the probe brane is
moving fast. As is clear from (\ref{n}), the time-like component of
the normal vector to the brane becomes important as the probe brane
enters into the fast-moving regime(see {\bf Fig. \ref{normalfig}} ).

At each point, we can span the 5-D space-time by the normal vector base
$\{e_{a} , n\}$, where $\{e_{a} \}$ are the tangent vector spanning
the brane world volume. More, specifically
\ba \label{ea}
e_{a}^{M} \, = \, \frac{\partial X^{M}}{\partial \xi^{a}} \,
\equiv \, X^{M}_{,a} \quad , \quad \ga_{ab} \, = \, G(e_{a}, e_{b}) \, ,
\ea
where $G(~,~)$ represents the inner product defined by the
background metric (\ref{metric1}), and $\gamma_{ab}$ is the induced
metric given in Eq. (\ref{gamma}). The normal vector $n^{M}$ is
defined by
\ba \label{n1} G(n, e_{a}) \, = \, 0 \quad , \quad  G(n,n) \, = \,
1, \ea
and the following projection relation holds
\ba \label{Projection}
\ga^{ab} X^{M}_{,a} X^{N}_{,b} \, = \, G^{MN} -n^{M}\, n^{N} \, .
\ea

For the background defined by Eq. (\ref{metric1}), the non-zero
components of the normal vector, $n^{0}$ and $n^{r}$, are
\ba
\label{n}
n^{0} \, = \, g^{1/2} h^{1/2} \dot R (1- h^{1/2} g \dot R ^{2}) ^{-1/2}
\quad, \quad
n^{r} \, = \, g^{-1/2}  (1- h^{1/2} g \dot R ^{2}) ^{-1/2} \, .
\ea

Using Eqs. (\ref{gamma}) and (\ref{C4}), the action of the moving brane
to all order in perturbations is
\ba \label{action2} S \, = \, -T_{3} \int d^{4} \xi \, \left(
\sqrt{-|\ga| } -  \frac{\partial\, X^{0}}{\partial\, t } C_{(4)}(X)
\right). \ea

The equation of motion for $\delta X^{M}$ which follows from this action
is obtained in {\bf Appendix A}, and is
\ba
\label{eom1}
\sqrt{-|\ga |} K \, = \, C_{(4)}' n^{r} \frac{\partial \,  X^{0}}{\partial \, t }- n^{0}  \frac{\partial\, C_{ (4) } }{\partial\, t} \, .
\ea
Here, $K$ is the trace of the extrinsic curvature of the surface,
$K= \ga^{ab}\, K_{ab}$. The extrinsic curvature is defined by
\ba
\label{Kab}
K_{ab} \, = \, e_{a}^{M}\, e_{b} ^{N} \, D_{M} \, n_{N}
\, = \, \frac{\partial X^{M}}{\partial \xi^{a}}
\frac{\partial X^{N}}{\partial \xi^{b}}  \,D_{M} \, n_{N}  \, ,
\ea
where $D_{M}$ is the five-dimensional covariant derivative which
is compatible with the metric $G_{MN}$.
One can check that at the homogeneous level Eq. (\ref{eom1})
reproduces the solution studied in the previous section.

To get the linearized equation of motion, we perturb Eq. (\ref{eom1})
about the background
(\ref{normal}). This is equivalent to expanding the action to second
order in the perturbations. We relegate the details of the calculations
leading to the perturbed equation of motion to {\bf Appendix A} and here just
give the final result
\ba
\label{eom2}
   \nabla^{2} \Phi -  M ^{2} \Phi  \, = \, 0 \, ,
\ea
where $\nabla^{2}$ is the Laplacian defined by the metric $\ga_{ab}$ and
the  effective mass term is given by
\ba
\label{M}
 M^{2} \, = \, - K_{ab} K^{ab} - R_{M N} n^{M} n^{N} + K^{2}
+\frac{1}{ \sqrt{-|\ga|}} ( n^{0} - \frac{n^{r} }{\dot R} )\,  (C_{(4)}' n^{r} )_{,t} \, .
\ea
In the above expression, $R_{MN}$ is the background Ricci tensor.
For the pure AdS background considered in \cite{Boehm:2002kf} only
the first two terms are present and the other two cancel out.

We now calculate the different contributions to $M^{2}$ for our background.
From Eq. (\ref{eom1}) at the homogeneous level, we find
\ba
\label{Kmass}
K \, = \, h\,  g^{-1/2}\,  C' = -\frac{h'}{h \,g^{1/2}}\, .
\ea
To calculate the other terms in $M^{2}$, we need to express $n^{0}$,
$n^{r}, \dot R$ and $\ddot R$ as functions of $r$ and $E$. To do this,
we use Eq. (\ref{eom0}) which allows $\dot R$ to be calculated.
Plugging the value of $\dot R$ obtained this way into Eq. (\ref{n}),
one can show that
\ba
\label{n0r}
n^{0} \, = \, E^{1/2}\, h^{3/4} \, (2+  E\, h )^{1/2} \quad \quad ,
\quad \quad
n^{r} \, = \, g^{-1/2}\, (1+ E h) \, .
\ea
Using these expressions for $n^{0}$ and $n^{r}$, then for the last term
in $M^{2}$ we obtain
\ba
\label{Cmass}
\frac{1}{ \sqrt{-|\ga|}} ( n^{0} - \frac{n^{r} }{\dot R} )\,  (C_{(4)}' n^{r} )_{,t} \, = \,
g^{-1/2} h \, \left( h' h^{-2}\, g^{-1/2}\, (1+ E\, h ) \right)' \, .
\ea

For the term $K^{ab} K_{ab}$, we have
\ba
\label{Kab2}
K^{ab} K_{ab} \, &=& \, \ga^{ac} \ga^{bd} K_{ab} K_{cd}  \nonumber\\
&=& \, \ga^{ac} \ga^{bd} X^{M}_{,a} X^{N}_{,b} X^{P}_{,c}X^{Q}_{,d}\, D_{M} n_{N}\, D_{P}\,n_{Q}
\nonumber\\
&=& \, D_{M} n_{N} \,D^{M} \,n^{N} - D_{M}\, n_{N}\, D_{P}\, n^{N}\, n^{M}\, n^{P} \nonumber\\
&=& \, \frac{h'^{2}}{4\, g \, h^{2}} (1+ 3 E^{2}\, h^{2} ) \, .
\ea
To go from the second to the third line, the orthogonality condition in
Eq. (\ref{n1}) and the projection relation Eq. (\ref{Projection}) have been
used.

Finally, for the contribution from the Ricci tensor, we get
\ba
\label{RicciM}
R_{M N} n^{M} n^{N}  =   \frac{-1}{8\,  g^{2} h^{2} }
 \left[  3 E h (2+ E h) ( 2 g h'^{2} -2 g h h'+ h h' g' )
 - 8 g h h'' + 4 h h' g'+ 10 g h'^{2} \right].
\ea

Plugging Eqs. (\ref{RicciM}), (\ref{Kab2}), (\ref{Cmass}) and (\ref{Kmass})
into the expression (\ref{M}) for $M^{2}$ we obtain
\ba
\label{finalM}
 M^{2} \, = \, \frac{E}{8 g^{2} h} \,
\left[ \, h\, (2+ 3 E\, h ) \, \left(g' h'-2 g h'' \right) + 4 g h'^{2}\, \right] \, .
\ea

\section{Mode Analysis}

The differential equation (\ref{eom2}) contains a Hubble friction
term for $\Phi$. As done in the usual theory of cosmological
perturbations (see \cite{MFB} for an in depth review and
\cite{RHBrev2} for a pedagogical overview), we can extract the
effects of the Hubble damping by the field redefinition
\ba
\label{phi}
\phi(\eta,\vec x) \, = \, a(\eta) \, \Phi(\eta,\vec x)
\, = \, h^{-1/4}\, \Phi \, ,
\ea
where for convenience we use conformal time $\eta$.
Eq. (\ref{eom2}) then becomes
\ba
\label{eom3}
\frac{\partial^{2} \phi}{ \partial \eta^{2}} -  \sum_{i} \partial_{i}^{2}\,  \phi + ( h^{-1/2} M^{2}
- \frac{a_{\eta \eta}}{a} )  \phi \, = \, 0 \, .
\ea

The above equation is analogous in structure to the equation which
describes the evolution of cosmological perturbations in cosmology
based on the usual four-dimensional Einstein gravity. On scales
larger than the Hubble radius, the negative square mass term
$a_{\eta \eta}/a$ is larger than the $k^2$ term coming from the
spatial gradients. Thus, whereas on sub-Hubble scales the gradient
term wins out and leads to the usual micro-physical oscillations,
the oscillations freeze out approximately at Hubble radius crossing,
and from then on the modes undergo squeezing. As is familiar from
the usual theory of cosmological perturbations, scalar metric
fluctuations are squeezed not with the factor $a(\eta)$ as would be
the case if $M^2 = 0$, but with a modified factor which is usually
called $z(\eta)$ which takes into account that $M^2 \neq 0$.

It is important to point out that, as shown in \cite{Boehm:2002kf},
for the observer on the brane our fluctuation $\Phi$ acts as the
usual Bardeen potential \cite{Bardeen} which is generally denoted by
$\Psi$. In terms of $\Psi$, in longitudinal gauge and in the
absence of anisotropic stress, the metric including
scalar metric fluctuations takes the form
\ba
ds^2 \, = - a^2(\eta)
\bigr[ (1 + 2 \Psi) d\eta^2 - (1 - 2 \Psi) d{\bf x}^2 \bigl] \, ,
\ea

 It is helpful to work in Fourier space where
\ba \phi_{k}(\eta) \, = \, \int d^{3} x \, \phi(\eta, \vec x) \, e^{
{-i \vec k . \vec x} }. \ea
This transforms Eq. (\ref{eom4}) into our desired mode equation
\ba
\label{eom4}
\frac{\partial^{2} \phi_{k}}{ \partial \eta^{2}} +(k^{2} -v_{eff} ) \, \phi_{k}  \, = \, 0 \, .
\ea
where the ``effective potential" is given by
\ba
\label{veff}
v_{eff} \, &=& \, \frac{a_{\eta \eta}}{a}  -h^{-1/2} M^{2}  \nonumber\\
&=&  \,  -h^{-1/2} M^{2} -\frac{h'}{ 4\, h}  \, \frac{\partial^{2 } R}{ \partial \, \eta^{2}}
-   (\frac{\partial\,  R}{ \partial \, \eta} )^{2}   \left( \frac{1}{4} \frac{h''}{h} - \frac{5}{16} \frac{h'^{2}}{h^{2}}
\right)\nonumber\\
&=& \, \frac{E^{2} h^{3/2}}{ 8\, g} \, \left( \frac{4 \, h''}{h} - 2 \frac{h'}{h}  \frac{g'}{g}  + \frac{h'^{2}}{h^{2}} 
\right),
\ea
where in the final line Eq. (\ref{reta}) is used to replace
$\partial \, R/\partial \, \eta$.

\begin{figure}[t]
\vspace{-2cm}
   \centering
   \hspace{-1.6cm}
   \includegraphics[width=4.1in]{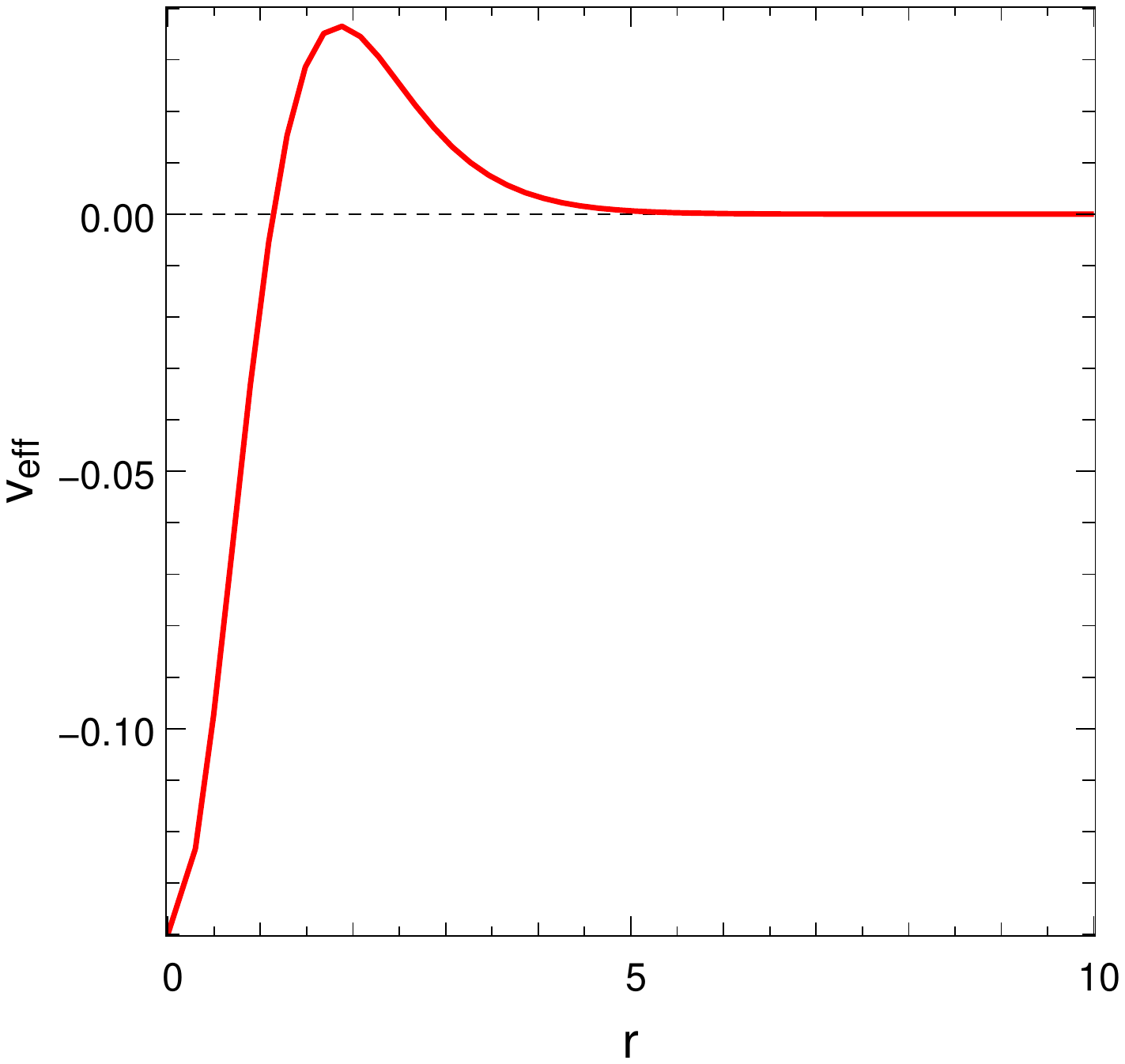} \hspace{-2.5cm}$
   \includegraphics[ width=4.1in]{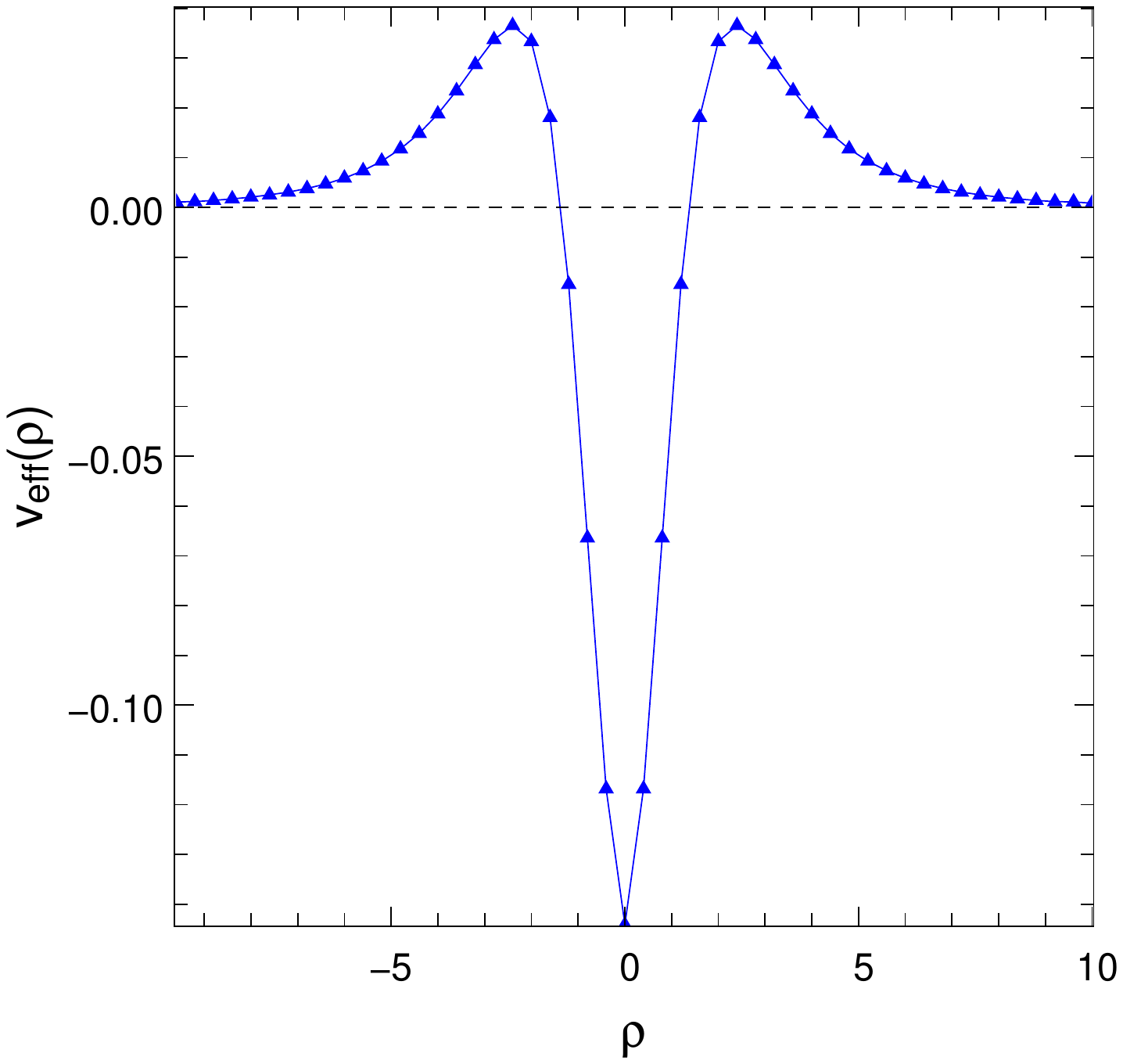} \hspace{-1.5cm} $
     \vspace{-5cm}
\caption{The plot on the left represents $\tilde v_{eff}$ as a
function of $r$. It has a global minimum at $r=0$ with $\tilde
v_{eff}\simeq -0.14$ and a maximum at $r\simeq 1.9$ with $\tilde
v_{eff}\simeq 0.04$ and exponentially falls off for large $r$. The
plot on the right represents $\tilde v_{eff}$ as a function of $\rho=(2E\beta)^{1/2}\tilde{\eta}$ .
The contracting and the expanding phases are mirror
images of each other. }
\label{potfig}
\vspace{0.8cm}
\end{figure}


We now make an observation which turns out to be important later on.
A closer look at (\ref{veff}) reveals that in the vicinity of $r=0$,
$v_{eff}$ is always negative. This is easily inferred by noting that
the 2nd and 3rd terms inside the bracket in (\ref{veff}) are both
vanishingly small around $r=0$ to ensure the smoothness of the
supergravity background. At the same time the fact that $h$ is a
monotonically decreasing function as a function of $r$ with a zero
first derivative at $r=0$ guarantees that $h^{''}/h$ is negative. So
one can conclude that around the bounce point, from the point of
view of the brane observer, the effective potential seen by the
brane fluctuation is negative.

We try to obtain some analytical insights into the evolution of the
mode functions given the effective potential of Eq. (\ref{veff}).
Due to the complexity of the KS geometry involving the functions
$I(r)$ and $K(r)$, it is not possible to have an analytical solution
for the entire range of $r$. However, in some limits we can perform
analytical calculations which can capture qualitatively (and even
quantitatively in the slow probe limit) the numerical results.

Let us define $h(r) = \beta \, I(r)$ and $g=\gamma\,
I(r)^{1/2}K(r)^{-2}$. Notice that Eq. (\ref{eom4}) can be rewritten
in the following form
\ba \label{eom5} \frac{d^2\phi_{k}}{d\tilde{\eta}^2} +
(\tilde{k}^2-\tilde{v}_{eff})\phi_{k} \, = \, 0, \ea
in terms of dimensionless variables
\ba \label{tilde} \tilde{\eta} \, \equiv \, \frac{E \beta^{3/4} }
{\gamma^{1/2} } \,  \eta \quad \quad , \quad \quad k \, = \,
\frac{E\beta^{3/4}}{\gamma^{1/2}} \, \tilde{k} \quad \quad , \quad
\quad v_{eff} \, = \, \frac{E^2\beta^{3/2}}{\gamma}\tilde
{v}_{eff}. \ea

The potential $\tilde{v}_{eff}$ is plotted in {\bf Fig. \ref{potfig}}. As a function of the
$r$-coordinate, it has a global minimum at the tip, $r=0$, reaches a
global maximum around $r\sim 1.9$, and falls off as $r \exp{(-2r)}$
for large values of $r$.  In the large $r$ limit, using the
asymptotic form of $I(r)$ and $K(r)$ given in {\bf Appendix B}, one
obtains
\ba
\label{veff_large_r}
\tilde v_{eff} \, \simeq \, \frac{20\times2^{1/3}}{6}\, re^{-2r} \, .
\ea

To solve the differential equation (\ref{eom5}) analytically,  we
need to express $\tilde{v}_{eff}$ as a function of $\tilde{\eta}$.
This requires inverting $r$ as a function of $\tilde{\eta}$. The
relation between $\tilde{\eta}$ and $r$ is given in Eq.
(\ref{reta}), which leads to
\ba
\label{invert}
\tilde{\eta} \, = \,
\pm \int_{0} ^{r} \frac{ dr\, }{K\, I^{1/2} \sqrt{1+ \frac{2}{ E \beta \, I} }}
\, .
\ea
The negative branch corresponds to the case when the brane is
falling towards the tip of the throat. $\tilde{\eta} = 0$
corresponds to the bounce point when the brane reaches the tip. The
positive branch corresponds to the case when the brane bounces back
towards the UV region of the throat. It is clear that
$\tilde{v}_{eff}$ is symmetric under $\tilde{\eta} \rightarrow -
\tilde{\eta}$.

The above integral can not be computed for all value of $r$.
However, in the limit of large values of $r$, when the KS solutions
is well-approximated by pure AdS geometry, one can compute the
integral analytically. Using the asymptotic forms of $I$ and $K$, we
have
\ba
\label{etatil}
\tilde{\eta} \, &\simeq& \, \pm  \,  2^{-1/6} \, 3^{-1/2}
\int^{r} dr'\, \frac{e^{r'}}{  \sqrt{ 1+ \frac{2^{4/3}}{3\, E \beta} \,  e^{4r'/3}   } } \nonumber\\
&=& \, \pm \,  2^{-1/6} \, 3^{-1/2} \,  e^{r} \, F( \frac{1}{2},
\frac{3}{4}; \frac{7}{4}, - \frac{2^{4/3}}{3\, E\, \beta} \,
e^{4r/3} ), \ea
where $F(a,b;c,z)$ is the hypergeometric function.

In two extreme regimes, Eq. (\ref{etatil}) can be inverted. First
consider the ``slow-roll'' limit when the brane is moving slowly and
$E\beta << 1$. This corresponds to the limit when the kinetic energy
of the brane is much smaller than the brane rest mass in the DBI
action in Eq. (\ref{Action}). Using the identity
\ba
\label{hypergeom}
 F(a,b;c,z) \, = \, (1-z)^{-a} F(a,c-b;c, \frac{z}{z-1}),
\ea
we obtain
\ba
\label{slowrolleta}
\tilde{\eta} \, \simeq \, \pm \, 3\times 2^{-5/6} \, e^{r/3} \, .
\ea

Using Eq. (\ref{veff_large_r}) for $\tilde{v}_{eff}$ in the large
$r$ limit, the potential reads
\ba
\label{slowrollv}
\tilde{v}_{eff} \, \simeq \,
\frac{C\,  \ln|  \tilde{\eta} | }{  {\tilde{\eta}}^{6} } \, ,
\ea
where $C$ is an un-important coefficient (the reason will be clear later).

The other limit where Eq. (\ref{etatil}) can be inverted is when
$E\, \beta \, e^{4r/3} >> 1$. This corresponds to the ``fast-roll''
limit, when the kinetic energy of the moving brane is comparable to
the brane rest-mass. In this limit, using the identity given in Eq.
(\ref{hypergeom}) for hypergeometric functions, we obtain
\ba \label{fastrolleta} \tilde{\eta} \, \simeq \, 2^{-1/6} \,
3^{-1/2}\,  e^{r}, \ea
which in the large $r$ limit leads to
\ba
\label{fastrollv}
\tilde{v}_{eff} \, \simeq \,
\frac{10}{9\, \ln | \tilde{\eta} |  } \, \frac{1}{ { \tilde{\eta} }^2 } \, .
\ea

It is interesting that in the fast-roll case, the
$\tilde{\eta}$-dependence of $\tilde{v}_{eff}$, up to a logarithmic
correction, has the same form as in the inflationary cosmology.
However, the constant prefactor is different: in the inflationary
model the prefactor is 2, while for the fast-roll case, it is
$10/9$. The constant has its origin in the asymptotic limit of the
KS geometry, where it reaches pure AdS space. As is well
known in the context of inflationary or Ekpyrotic cosmology, the
constant 2 is crucial to obtain a scale-invariant spectrum starting
with the Bunch-Davis vacuum initial conditions. This change in the
coefficient results in a non-scale-invariant prediction for the
scalar spectral index $n_{s}$, as we will see explicitly in the next
section.

The effective potential $\tilde{v}_{eff}$ as a function of
$\rho=(2E\beta)^{1/2}\tilde{\eta}$ is given in {\bf Fig. \ref{potfig}}. Suppose
the brane starts from the UV region heading towards the IR region of
the throat, so $\tilde{\eta} <0$ initially. The modes start in the
sub-Hubble region where $-\tilde{k}  \, \tilde{\eta} >> 1$ and
${\tilde{k}} { \, ^2}-\tilde{v}_{eff} > 0$. They evolve as
sub-horizon mode until ``crossing the potential'' (i.e. ${\tilde{k}}
{ \, ^2} - \tilde{v}_{eff} = 0$) and then become super-Hubble. They
remain super-Hubble until they cross the potential again near $r =
0$ and become sub-Hubble. The evolution of the modes is sub-Hubble
near the tip of the throat until the brane bounces back at $r=0$ or
$\tilde{\eta}=0$. The mode evolution for $\tilde{\eta}>0$ is the
mirror image of what happens for $\tilde{\eta}<0$.

The scalar spectral index can be calculated at an arbitrary point
like $\tilde{\eta_{*}}$, as long as the mode is super-Hubble but in
the decelerating phase at late times. One may identify $\tilde{\eta}
= \tilde{\eta_{*}}$ with the surface of last scattering.

\section{Solving the Mode Equation Analytically}

Having obtained the effective potential $\tilde{v}_{eff}$ in the large
$r$ limit of both fast-roll and slow-roll cases, we can solve the mode
equation and compute the spectrum of fluctuations using analytical
approximations. First, recall that
the scalar power spectral index $n_s$ is given by
\ba
\label{power}
P_{\phi}(k) \, \equiv \, \frac{k^{3}}{ 2 \pi^{2}} \, |  \phi_{k} |^{2}
\, \sim \, k^{n_{s}-1} \, .
\ea

Let us start with the fast-roll case which is more amenable to
analytic approximation and where the equations are more similar
to those appearing in inflationary models.
To solve the differential equation (\ref{eom5}) analytically, we make a crude
approximation here and neglect the logarithmic running in the
effective potential (\ref{fastrollv}). We simply replace the logarithm by
one. Comparing with the exact result obtained numerically in the next section,
we find that this is a reasonable approximation.

We wish to compute the power spectrum of $\phi_{k}$ at a late time
$\eta_{*}$ on super-Hubble scales. Let us denote the time that the
mode $k$ crosses the Hubble radius shortly after the bounce by
$\eta_2(k)$, the time it enters the Hubble radius shortly before the
bounce by $\eta_1(k)$, and the time it exits the Hubble radius
during the initial contracting phase by $\eta_{0}(k)$. Due to the
fact that the potential is very negative at the bounce point, in
small $k$ limit, the k-dependence of $\eta_2(k)$ and $\eta_1(k)$ is
negligible. It is crucial to note that the negativity of the
potential in the vicinity of the bounce point is generic as it was
argued in the previous sections. Also, between $\eta_1(k)$ and
$\eta_2(k)$, i.e. in the bounce region, the mode is oscillating and
its amplitude does not change. This implies that as long as
the potential around the bounce point is negative, the bounce region
does not introduce any k-dependence in the transfer matrix.
Hence, the power spectrum of $\phi_{k}$ at the late time $\eta_{*}$
has the same spectral index as the power spectrum at pre-bounce time
$-\eta_{*}$:
\ba
P_{\phi}(k, \eta_{*}) \, \equiv \, k^3 |\phi_{k} (\eta_{*}) |^2
\, = \, k^3 {\cal A} \,  |\phi_{k}(-\eta_{*}) |^2 \, ,
\ea
where the amplification factor ${\cal A}$ is given by
\ba
{\cal A} \, = \, \left( {|\phi_{k} (\eta_{*}) |^2} \over {|\phi_{k} (\eta_2(k))  |^2 } \right)
 \left( {   |\phi_{k} (\eta_2(k))  |^2   } \over {  |\phi_{k} (\eta_1(k))  |^2    } \right)
  \left( {   |\phi_{k} (\eta_1(k))  |^2   } \over {  |\phi_{k} (-\eta_{*}(k))  |^2    } \right)
\ea
and is, to a first approximation, independent of $k$. Thus, the spectral
shape is determined by the growth of the mode functions between initial
Hubble radius crossing at $\eta_{0}(k)$ and the time $-\eta_{*}$:
\ba \label{spectrum}
P_{\phi}(k, \eta_{*}) \, \sim \,
k^3 {{|\phi_{k} (-\eta_{*})  |^2 } \over {|\phi_{k} (\eta_{0}) |^2}}
|\phi_{k}  (\eta_{0}) |^2 \, .
\ea

In the fast-roll case, the dominant solution of $\phi_{k}$ in the contracting
phase is given by
\ba \label{scaling}
\phi_{k}(\eta) \, \sim \, \eta^{-2/3} \, .
\ea
Since $\eta_{0}(k)$ is given by $\tilde k^2 = \tilde
v_{eff}(\eta_{0}(k))$ we have
\ba \label{Hubbleexit}
\eta_{0}(k) \, \sim \, k^{-1} \, .
\ea
Inserting (\ref{scaling}) and (\ref{Hubbleexit}) into (\ref{spectrum})
and assuming that the modes start out on sub-Hubble scales during
the period of contraction (i.e. for $\eta < \eta_{0}(k)$) in
the vacuum state with the Bunch-Davis initial condition
\ba \label{BD} \phi_{k} \sim \frac{A}{ \sqrt{2 k}} \,  e^{ - i\,
\tilde k \,  \tilde{\eta} } \,, \ea
we obtain
\ba
P_{\phi}(k) \, \sim \, k^{2/3} \, ,
\ea
and thus the spectral index is predicted to be $n_s = 5/3$.

To be slightly more precise, using the large $r$ limit of $\tilde
v_{eff}$ form Eq. (\ref{fastrollv}), the solution of (\ref{eom5}) is
given in terms of Hankel functions of first and second kind
\ba \phi_{k} \, = \, \sqrt{\tilde{\eta}} \,\left[ A_{1} \,  {H_{7/6
} } ^{(1)} (-\tilde{k}\, \tilde{\eta}) +   A_{2} \,  {H_{
7/6}}^{(2)} (-\tilde{k}\, \tilde{\eta}) \right ], \ea
where $A_{1}$ and $A_{2}$ are two constants of integration.
Like in inflationary models, only the Hankel function of first type
matches with the initial vacuum condition since
\ba
 {H_{ 7/6}}^{(1)} (x>>1) \, \sim \, \sqrt { \frac{2}{\pi x}}
e^{ \, i (x- \frac{10\, \pi}{12 } )} \quad \quad , \quad \quad
 {H_{ 7/6}}^{(2)} (x>>1) \, \sim \, \sqrt { \frac{2}{\pi x}}
e^{\, - i (x- \frac{10\, \pi}{12 } )}, \ea
which implies that $A_{2}=0$.
Thus, we obtain the scaling of $\phi_{k}$ with $\eta$ given in
(\ref{scaling}), and, as shown above, this leads to
\ba
\label{fastrolln}
n_{s} \, = \, 5/3 \, .
\ea
Interestingly enough, the spectrum is not scale invariant. As
explained before, this is due to the numerical factor $10/9$ in Eq.
(\ref{fastrollv}) for $\tilde{v}_{eff }$. This particular number
originated from the fact that the UV region of the throat is
well-approximated by pure AdS geometry. This indicates that for pure
AdS geometry, the spectral index will be $n_{s} = 5/3$. As explained
before, we have neglected the logarithmic correction into
$\tilde{v}_{eff }$, which originates from the logarithmic correction
to the AdS geometry in the KS solution. Considering the logarithmic
correction, our numerical analysis (presented in next section) show
that $n_{s}\simeq 2.3$.

The case of slow-roll proved to be more involved. Even when we
neglect the logarithmic correction for $\tilde{v}_{eff }$ in Eq.
(\ref{slowrollv}), we could not solve the differential equation
(\ref{eom5}) simultaneously for both sub-Hubble and super-Hubble
modes. However, we can solve it for sub-Hubble and super-Hubble
modes separately. For sub-Hubble modes the solution is as given by
vacuum state Eq. (\ref{BD}). For super-Hubble modes , the solution
of Eq. (\ref{eom5}) with effective potential Eq. (\ref{slowrollv})
is given in terms of Bessel and Neumann functions
\ba
\label{slowrollsup}
\phi_{k} \, = \, \sqrt{\tilde{\eta} } \left[ \tilde B_{1} \, J_{-1/4}\,  (  \frac{\sqrt{-C}   }{ 2 \tilde{\eta}^{2}  } )
+  \tilde B_{2}  \, Y_{-1/4}\,  (  \frac{\sqrt{-C}   }{ 2 \tilde{\eta}^{2}  }  )
\right] \, .
\ea
At the transition $\tilde{\eta}_{0}(k)$ point when ${\tilde{k}} { \,
^2}-\tilde{v}_{eff} = 0$, this solution should smoothly go over to
the solution for the sub-Hubble modes. The transition point is given
by
\ba \tilde{\eta}_{0}(k) \, =  C^{1/2} \, k^{-1/3}. \ea
Since we are interested in large scale perturbations, $k<<1$, so
$\tilde{\eta}_{0}(k) >> 1$ and we can use
the small argument approximation for the Bessel and Neumann functions
\ba
 J_{\nu} (x<<1) \, \sim \, x^{\nu} \quad \quad , \quad \quad
 Y_{\nu} (x<<1) \, \sim \, - x^{-\nu} \, .
\ea
Using these approximations in Eq. (\ref{slowrollsup}), we obtain
\ba
\label{phi0}
\phi_{k}( \tilde{\eta}_{0}  ) \, \simeq \, (B_{1} + B_{2} \, \tilde{\eta}_{0} ) \, ,
\ea
where $B_{1}$ and $B_{2}$ are new constant of integrations related to
$\tilde B_{1}$ and $\tilde B_{2}$. Matching this solution to the
sub-Hubble solution Eq. (\ref{BD}) at $\tilde{\eta}_{0}(k)$, we find
\ba
B_{1} \, \simeq \, \frac{  A}{\sqrt{2\, k}}(1+ i\, k\,
\tilde{\eta}_{0}(k) ) \,   e^{-i \, k\,  \tilde{\eta}_{0}(k)  }
\quad \quad , \quad \quad
B_{2} \, \simeq \, \frac{-i\,  A}{\sqrt{2}} \sqrt{k} \, e^{-i
\, k\,  \tilde{\eta}_{0}(k) } \, .
\ea
For large scale perturbations where
$k \rightarrow 0$, $B_{1}$ scales like $k^{-1/2}$, while $B_{2}$ scales
like $k^{1/2}$ and from  Eq (\ref{phi0}) we obtain
$\phi_{k}( \tilde{\eta}_{0}  ) \, \sim k^{-1/2}$ .
Using this in Eq (\ref{spectrum}), one obtains
\ba
n_{s} \, = \, 3 \, .
\ea
We have calculated the spectral index numerically and found
$n_{s}=3\pm 10^{-5}$, in agreement with the above analytical value.

Indeed, one can argue that for potentials of the form ${\tilde
v}_{eff} \sim \tilde{\eta}^{-n}$, with $n > 2$, the spectral index
is always equal to $3$. On the other hand, $n=2$ is the critical
value where the spectral index crucially depends on the numerical
coefficient in ${\tilde v}_{eff}$. For a potential of the form
${\tilde v}_{eff} = c \, \tilde{\eta}^{-2}$, one can show that
$n_{s}= 4- \sqrt{1+4c}$. For inflationary models in the limit where
slow-roll corrections are ignored $c=2$ and one obtains $n_{s}=1$.
In our fast-roll case $c= 10/9$, which leads to $n_{s}=\frac{5}{3}$,
as we have shown before.

\section{Numerics}

\begin{figure}[t]
\vspace{-2cm}
   \centering
   \hspace{-1.6cm}
   \includegraphics[width=4.1in]{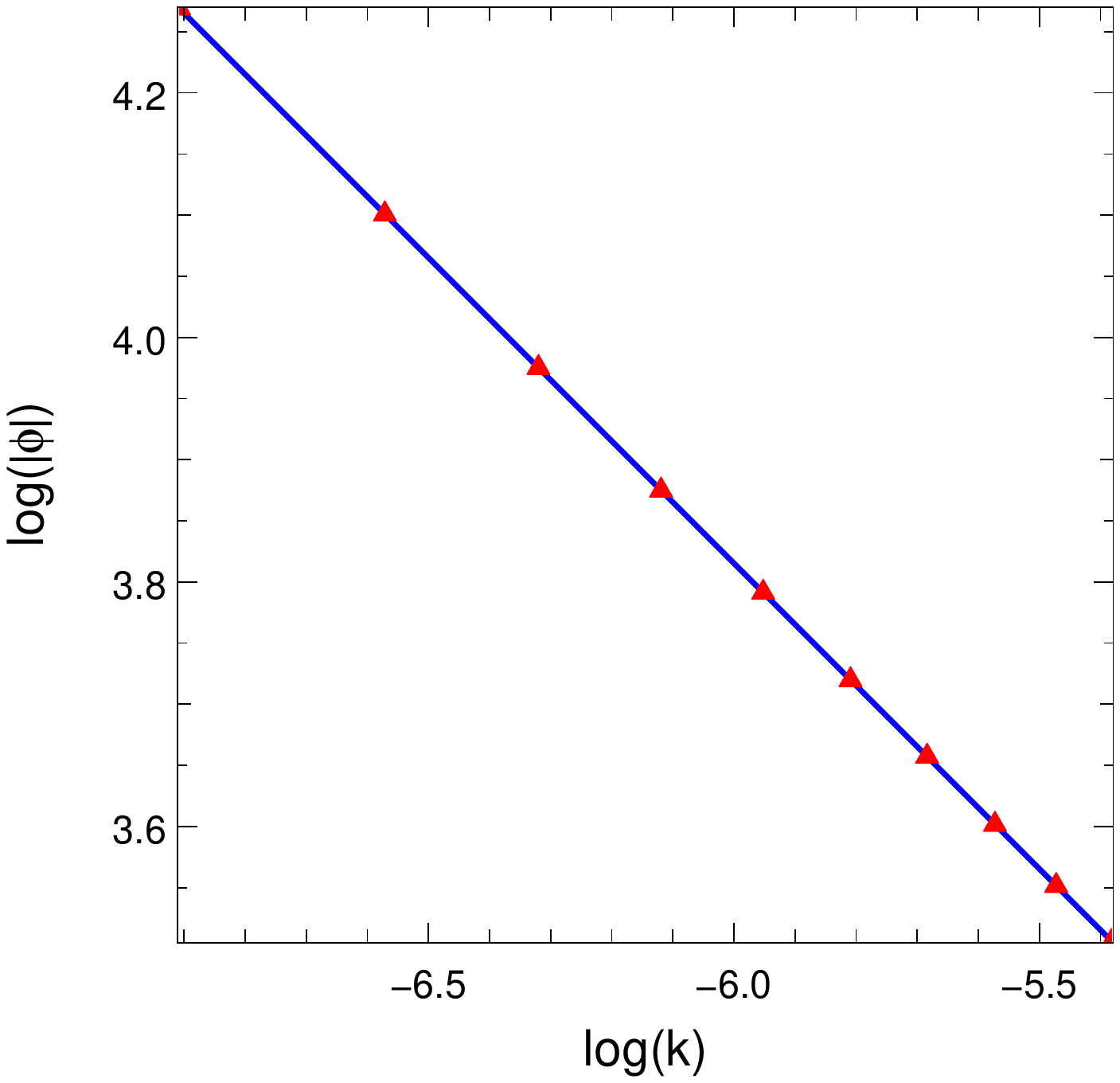} \hspace{-3.5cm}$
   \includegraphics[ width=4.1in]{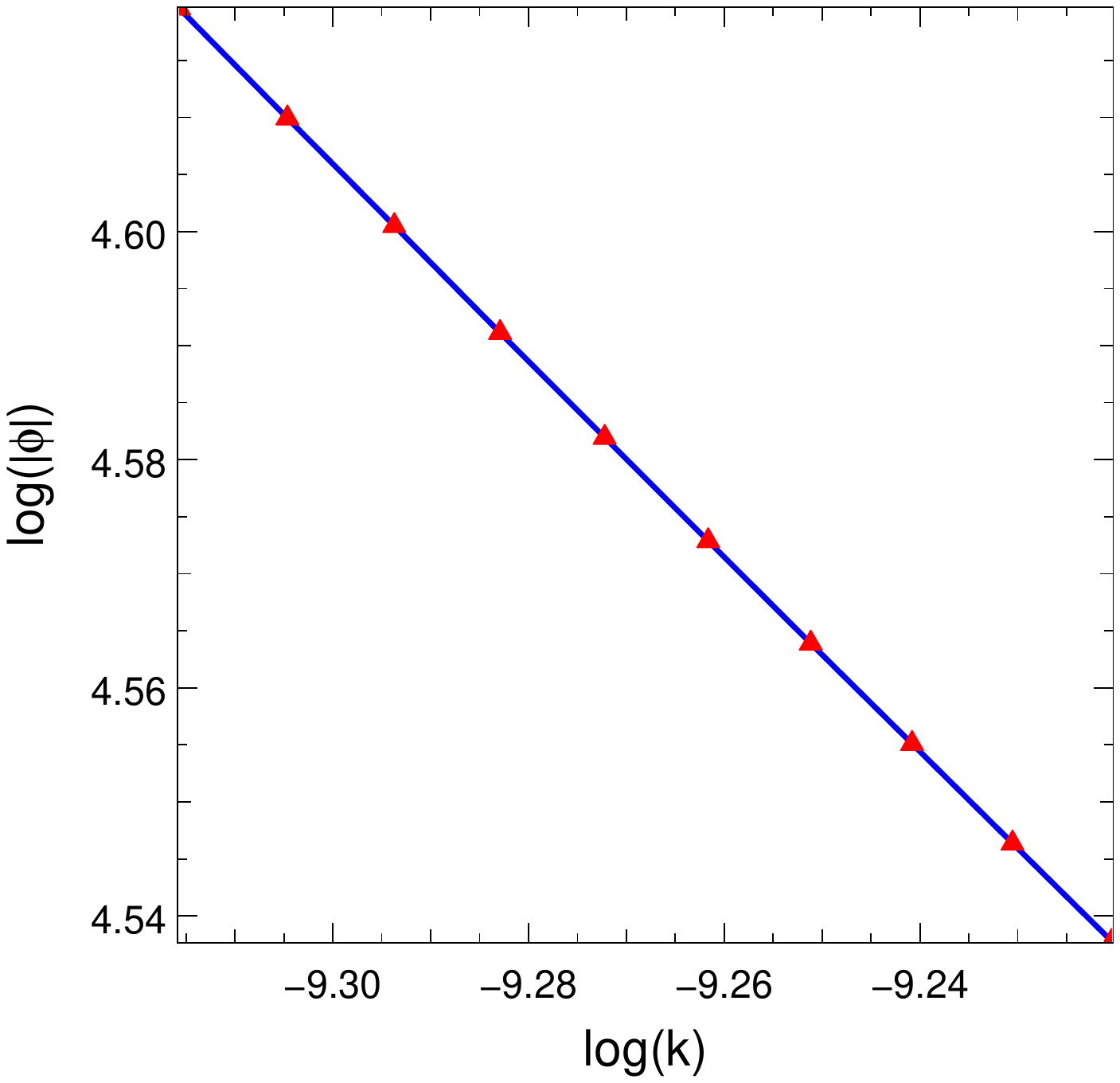} \hspace{-1cm} $
   \vspace{-5cm}
\caption{The $k$-dependence of $\phi_{k}(\eta_{*})$ is numerically plotted.
The figure on the left corresponds to the slow-roll limit with the slope equal to $-0.5$ and 
$n_{s}=3$, while that of the right hand side corresponds to the fast-roll case with slope equal to 
$-0.85$ and $n_{s}=2.3$. The data points are shown by red triangles.}
\label{specplot}
\vspace{0.8cm}
\end{figure}

In this section we discuss our numerical results. In order to solve
the equation of motion for $\phi_{k}$ (\ref{eom5}) numerically, we find
it more convenient to work with the coordinate $r$ related to $\tilde \eta$
through (\ref{invert}). In this coordinate, we look for a left
moving wave solution initialized at very large values of $r$ which
eventually reflects back to the large $r$ region. The equation of
motion is initialized with the Bunch-Davis vacuum initial data given
in Eq (\ref{BD}).
Boundary conditions at the origin of the $r$-coordinate are easily
found to be the impact boundary conditions
\begin{eqnarray}
\phi_{k \, in}|_{r=0^{+}} \, &=& \, \phi_{k\, out}|_{r=0^{-}},\\ \nonumber
\frac{d\phi_{ k \, in}}{dr}|_{r=0^{+}}\, &=&-\, \frac{d\phi_{k\, out}}{dr}|_{r=0^{-}} \, ,
\end{eqnarray}
where $\phi_{k\, in}$ and $\phi_{ k\, out}$ refer to the incoming
and outgoing waves at the origin, respectively. The quantity of
interest to us is the spectral index defined by (\ref{power}). As
usual, $\phi_{k}$ is evaluated at a convenient point in the
super-Hubble region. It is worth emphasizing that picking a
different point will only amount to a change in the amplitude of the
power spectrum and will not alter its k-dependence or the index as
long as it lies within the super-Hubble region. We solve
(\ref{eom5}) for different values of $k$ and read off the index from
the log-log plot of $k^3|\phi_{k}|^2$ versus $k$, generated
numerically for a range of $k$ values. We observe a negligible
running of the index, for either small or large values of the
parameter $E\beta$, as we dial $k$. We made sure that a change in
the location where the initial Bunch-Davis vacuum condition was
imposed on our numerical solution did not influence the resulting
index. {\bf Fig. \ref{specplot}} illustrates our results for the index in two
different limits for $E\beta$. The left side represents the results
in the small $E\beta$ limit. The right side gives the result in the
opposite limit where $E\beta$ is large (by which we simply mean the
limit where the second term under the square root in (\ref{invert})
can be ignored). Note that the limit itself depends on the
wavelength of the observed large scale structures, $k_{obs}^{-1}$,
around which the index is calculated.


\begin{figure}[t]
\vspace{-2.5cm}
   \centering
   \hspace{-1.6cm}
   \includegraphics[width=4.4in]{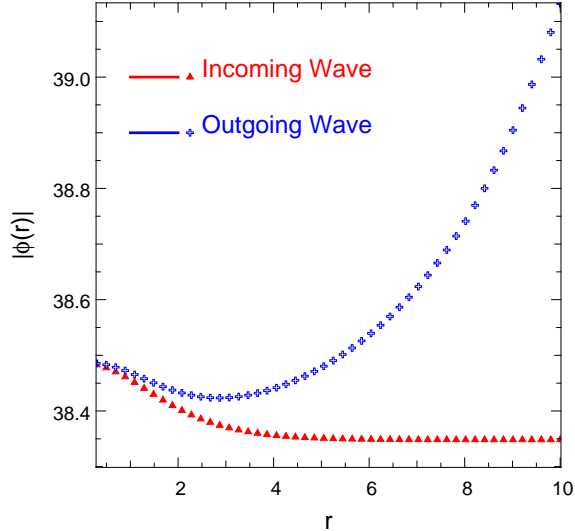}
   \vspace{-5cm}
  \caption{The wave function is presented here. The red(blue) curve corresponds to
  contracting(expanding) phase. As  time increase, i.e. $\tilde \eta$ increases, the wave function also increases. This indicates that the dominant mode in the contracting phase is matched to the dominant mode in the expanding phase.}
  \label{waveplot}
\vspace{0.8cm}
\end{figure}

{\bf Fig. \ref{waveplot}} shows a sample wavefunction (its modulus to be exact)
plotted for $\tilde{k}=0.0034$. The lower curve represents the
wavefunction in the contracting phase. The amplitude is growing as
the bounce point $r = 0$ is approached. The upper curve is
the amplitude of the wavefunction in the expanding phase. As is
apparent, the amplitude is growing in time (i.e. as $\tilde{\eta}$ increases).
This demonstrates explicitly that the dominant mode of $\phi_k$ in
the contracting phase couples with un-suppressed amplitude
to the dominant mode in the expanding phase, unlike what is obtained
in models in which a contracting Einstein cosmology is matched to an
expanding Einstein cosmology at a singular hypersurface making use
of the usual matching conditions of \cite{Hwang,Deruelle}.

Note that in the simulation of {\bf Fig. 4}, the initial condition
was imposed at $r=10$, where the brane is deep in the UV region,
well-approximated by the AdS geometry.


\section{backreaction}

As discussed in \cite{Kachru:2002kx}, we need to check that the
gravitational instability associated
with long wavelength gravitational perturbations does not destabilize the
homogeneous background.
The length scale associated with this instability is the Jeans length
given by $L_{J}= v_{s}/\sqrt{G\, \rho}$
where $v_{s}$ is the speed of sound,
$G$ is the Newton's constant and $\rho$ is the energy density on the brane.
The time scale associated with this instability is $t_{ins} \geq L_{J}$.

For our model, from Eq. (\ref{Action}), we obtain
\ba
\label{rho}
\rho \, &=& \, T_{3} h^{-1} \, (1- \sqrt{1- g h^{1/2}{ \dot r}^{2}  }) \nonumber\\
&=& \, \frac{E\, T_{3}}{ 1+ E\, h} \, .
\ea
In the slow-roll limit, one can see that $\rho \simeq E T_{3}$ is constant.

Using $G \leq g_{s}^{2} l_{s}^{2}$ for our compactification \cite{Kachru:2002kx}, we have
\ba
t_{ins} \, \geq \, L_{J} \, = \,
\frac{v_{s}}{\sqrt{G \rho}} \geq v_{s} l_{s} \sqrt{\frac{1+E\, h} { E} } \, .
\ea

Suppose one releases the brane from the point $r=r_{*}$ with the
initial velocity $v_{*}$. Since the brane is accelerating towards
the tip of the throat, the time for the bounce satisfies
\ba t_{bounce} \, \leq \, \frac{2 d_{*}}{ v_{*}}, \ea
where $d_{*}$ is the physical distance traveled by the brane before
reaching the bottom of the throat. One can show that
\ba
v_{*} \, = \, \sqrt{ E h(r_{*})  (2 + E\, h(r_{*}) )}
\ea
Discarding factors of order unity, taking $v_{s} \sim 1$ and using the fact that 
$\epsilon ^{4/3} e^{2r_{*}/3} \sim l_{s}^{2}$
 one can show
\ba
\frac{t_{bounce}}{t_{ins}} \, < \,
\frac{r_{*}^{3/4} }{\sqrt{g_{s} M}} \, { \left[    (1+E h(r_{*}) )
(2 + E\, h(r_{*}) )\, \right]}^{-1/2} \, .
\ea

In the slow-roll limit when $E h(r_{*}) << 1$, one reaches the bound
obtained in \cite{Kachru:2002kx}. We see that this bound gets even
weaker for the fast-roll limit, when $E h(r_{*}) >> 1$. As an
estimate, one can take $r_{*} \sim  (5-10)$, when the brane is deep
inside the throat. Also, to trust the supergravity limit, we take
$g_{s} M >> 1$. In this limit it is thus justified to neglect the
effect of gravitational instability on the brane.


\section{Discussion}

We have studied the evolution of cosmological fluctuations in a mirage
cosmology setup in which the observer lives on a BPS D3-brane which
is moving into and out of a Klebanov-Strassler throat. This provides
a simple model of a non-singular bouncing cosmology.

Our main results are twofold. First, we find that the growing mode
of the cosmological fluctuations in the contracting phase couples
without suppression to the growing mode in the expanding phase. This
is unlike what happens in a setup in which the fluctuations are
treated in Einstein gravity in both the expanding and contracting
phase, and then matched through a singular hypersurface using the
analog of the Israel matching conditions. Negativity of the
effective potential felt by the perturbations in the neighborhood of
the bounce point appears to be generic in the models of the kind
studied in this paper. As we saw, this leads to the fact that the
bounce region does not contribute any $k$-dependence to the spectral
index. Secondly, we find that if we set off the modes in their
Bunch-Davies vacuum on sub-Hubble scales in the contracting phase, a
final spectrum in the expanding phase which is not consistent with
observations will emerge. The specific spectral index depends on
whether the brane is moving fast or slow. In the latter case, the
resulting spectral index is $n_s = 3$, in the former $n_s$ is closer
to $2$ but still completely inconsistent with the data which demands
$n_s = 0.95 \pm 0.05$.

In this analysis we did not take into account the effects of volume modulus
stabilization, such as can be achieved by wrapped D7-branes, on the
mobile D3-brane. In a realistic model where all back-reactions are included,
the brane feel an attractive force towards the tip of the throat
\cite{Baumann:2006th, Burgess:2006cb}. This changes $v_{eff}$ and  it is
interesting to see whether or not these effects can modify the spectral index.
It is also possible that mirage cosmology motion in a different type of throat
might lead to a spectrum consistent with the cosmological data.
It is also possible that a pre-cursor phase which involves extra physics
(as assumed e.g. in the Ekpyrotic scenario) leads to a consistent spectrum.
Finally, it is possible that, if the bounce phase lasts sufficiently long,
thermal fluctuations of a gas of closed strings with winding modes will
generate a scale-invariant spectrum \cite{NBV,BNPV2,RHBrev3}.


\begin{acknowledgments}{We thank D. Easson, C. Germani, N. Grandi, J. Khoury, B. Ovrut,
D. Steer and A. Tolley for useful discussions and K. Hassani
for computer assistance. This work is supported by NSERC under the
Discovery Grant program, by Canada Research Chair funds (RB), and by
funds from a FQRNT Team Grant. O.S. is supported in part by a McGill
Tomlinson Postdoctoral Fellowship. }\\
\end{acknowledgments}


\appendix

\section{Linear equation of Motion}

Here we present the details of the calculations leading to the linear
perturbation equation (\ref{eom2}).
Although we perform the analysis for a D3-brane moving in a five
dimensional background but
the result is applicable to a Dp-brane with co-dimension
one, i.e. $p=D-1$, where $D$ is the background space-time dimension.

As was briefly described in Section 2, at each point in space-time we can
choose the normal vector base
$\{ e_{a} , n\}$, such that
\ba
e_{a}^{M} \, = \, \frac{\partial X^{M}}{\partial \xi^{a}}  \equiv X^{M}_{, a}
 \quad , \quad
\ga_{ab} \, = \, G(e_{a}, e_{b})
\ea
and
\ba
G(n, e_{a}) \, = \, 0 \quad , \quad G(n, n) \, = \, 1 \, ,
\ea
where $G(u,v)\equiv G_{M N} u^{M} v^{N}$.

The following projection relation also holds:
\ba
\label{projection}
\ga^{ab} X^{M}_{,a}\, X^{N}_{, b} \, = \, g^{MN}- n^{M}\,n^{N} \, .
\ea

We are interested in the normal displacements of the brane
\ba
\label{normal2}
X^{M} \, = \, \bar {X }^{M} + \Phi\, n^{M}  \, .
\ea

Under this transformation, we have
\ba
\label{deltagamma}
\delta \ga_{ab} \, &=& \, \Phi \left(\,  G_{MN, P}\, n^{P}\,  X^{M}_{,a}\, X^{N}_{,b} + 2 G_{MN}\, n^{M}_{,a}\, X^{N}_{,b} \right )\nonumber\\\
&=& \, 2 K_{ab} \Phi
\ea
where $K_{ab}$ is the extrinsic curvature of the surface defined in
Eq. (\ref{Kab} ).

The equation of motion for the normal displacement $\Phi$ from the
action (\ref{action2}) is obtained by
\ba
\delta_{\Phi} S&=& -T_{p}  \, \int d^{p} \xi \, \left(  \frac{1}{2}\sqrt{-|\ga| } \gamma^{ab} \delta \gamma_{ab}
-  \frac{\partial\, X^{0}}{\partial\, t }\, C_{(4),M} \, \delta X^{M}- C_{(4)} \, \frac{\partial\, \delta X^{0}}{\partial\, t }
\right)\nonumber\\
&=& -T_{p}  \, \int d^{p} \xi \,  \left[ \sqrt{-|\ga| } \,  K -
\frac{\partial\, X^{0}}{\partial\, t } \, C_{(4),M}\, n^{M} + n^{0}
\frac{\partial\, C_{(4)}}{\partial\, t }  \right] \, \Phi \, , \ea
which produces Eq. (\ref{eom1}).

To get the linearized equation of motion we perturb Eq. (\ref{eom1})
around Eq. (\ref{normal2}). The perturbed equation of motion is to
some extent involved. Here we present the outline of the
calculations. The linear equation is obtained by perturbing the left
hand side (LHS) and right hand side (RHS) of Eq. (\ref{eom1})
separately.

For the LHS, we have
\ba
\label{LHS1}
\delta(LHS) \, = \, \delta( \sqrt{-|\ga| }  K )\, = \,
\sqrt{-|\ga| } \left [( K^{2} - 2 K^{ab} K_{ab} )\Phi
 + \ga^{ab}   \delta K_{ab}  \right] \, .
\ea
To obtain the terms proportional to $\Phi$, the relation (\ref{deltagamma})
for $\delta \ga_{ab}$ was used. To calculate $\delta K_{ab}$, it is very useful
to perform the analysis in locally Gaussian normal coordinates (GNC), where
at each point
$G_{MN, P} = \Gamma^{P}_{QS}=0$, where $\Gamma^{P}_{QS}$ is the connection
compatible with the metric $G_{MN}$.
Starting with the expression for $K_{ab}$ given in Eq. (\ref{deltagamma}),
in  GNC we have
\ba
\label{deltaKab}
\ga^{ab}\delta K_{ab} \, = \,
\frac{\Phi}{2} n^{P} n^{Q}\, ( G^{M N} -n^{M} n^{N} ) G_{MN,PQ} +
\ga^{ab} G_{MN} (\delta X^{M}_{,b} n^{M}_{,a} +X^{N}_{,b} \delta n^{M}_{,a} )
\ea
where to obtain the terms proportional to $\Phi$ the projection identity
Eq. (\ref{projection}) was used.
We now calculate the last two terms in Eq. (\ref{deltaKab}) separately.
To do that we note that
\ba
\label{deltan}
\delta n^{M} \, = \, A^{a} e_{a}^{M} + B\, n^{M}
\ea
where
\ba
B \, &=& \, -\frac{\Phi}{2} \, G_{MN, P} \, n^{M} n^{N} n^{P} \\
A^{a} \, &=& \, - \ga^{ab} \Phi_{,b} - \ga^{ab} \,
(G_{MN, P}\, X^{N}_{,b} \, n^{M}n^{P} + G_{MN} \, n^{M} n^{N}_{,b} ) \Phi \, .
\nonumber
\ea

Using this for the third term in Eq. (\ref{deltaKab})  we obtain \ba
\label{secterm} \ga^{ab} G_{MN}  \delta X^{M}_{,b} n^{M}_{,a} \, &=&
\,
\ga^{ab} G_{MN} n^{M}_{,a} ( n^{N}_{,b} \Phi + n^{N} \Phi_{,b}) \nonumber\\
 &=& \, \ga^{ab} G_{MN} n^{M}_{,a}  n^{N}_{,b} \Phi \nonumber \\
&=& \, \Phi K^{ab}K_{ab} \, .
\ea
To go from the first line to the second line, the relation
$n^{M}n_{M,a} = 0$ which holds in GNC was used.
The final identity in the above equation is easy to prove in GNC
following a chain identity for partial derivatives.

Similarly for the last term in Eq. (\ref{deltaKab})  we have
\ba \label{thirdrerm}
\ga^{ab} G_{MN}  X^{N}_{,b} \delta n^{M}_{,a} \, &=& \, \ga^{ab} G_{MN}
( X^{M}_{,c} X^{N}_{,b} \, A^{c}_{,b} + X^{N}_{,b}\,X^{M}_{,ac} \, A^{c})
\nonumber\\
&=& \, A^{a}_{,a} + A^{c} \ga^{ab} G_{MN} X^{N}_{,b} X^{M}_{,ac} \, .
\ea
One notes that the term containing $B$ vanishes in GNC while the term
containing $B_{,a}$ vanishes due to relation $G(n, e_{a})=0$.

After some algebra, and working in GNC, one can show that
\ba
A^{a}_{,a}  &=& - \left[( G^{P N} -n^{P} n^{N} ) \, n^{M} n^{Q}  G_{MN,PQ} +
\ga^{ac} G_{MN} n^{M}n^{N}_{,ac} +  K^{ab}K_{ab} \right] \Phi\nonumber\\
&-& \ga^{ac} \Phi_{,ac} - \ga^{ac}_{,c} \Phi_{,a}
\ea
and
\ba
A^{c} \ga^{ab} G_{MN} X^{N}_{,b} \, X^{M}_{,ac} \, = \, - \ga^{ab} \ga^{cd}
G_{MN} X^{N}_{, b} X^{M}_{,ac} \Phi_{,d} \, .
\ea

Combining all the terms, we obtain
\ba \label{thirdterm2}
\ga^{ab} G_{MN}  X^{N}_{,b} \delta n^{M}_{,a} \,
&=& \, - \nabla^{2} \Phi - K^{ab}K_{ab} \Phi - \ga^{ac}\, G_{MN} n^{M}n^{N}_{,ac}  \Phi \nonumber\\
& & \, - ( G^{P N} -n^{P} n^{N} ) \, n^{M} n^{Q}  G_{MN,PQ} \, \Phi
\ea
where $\nabla^{2}$ is the Laplacian defined by the metric $\ga_{ab}$.

Combining Eqs. (\ref{thirdterm2}), (\ref{secterm}) and
(\ref{deltaKab}) we obtain
\ba \label{deltaKab2}
\ga^{ab}\delta K_{ab} \, &=& \, - \nabla^{2} \Phi +
\frac{1}{2} n^{P} n^{Q} g^{MN}\, ( g_{M N, PQ} -2 g_{NP,MQ} )  \Phi \nonumber\\
& & \, + n^{M} n^{N} n^{P} n^{Q} g_{PQ, MN}  \Phi
 -   \ga^{ab} G_{MN} n^{M} n^{N}_{, ab}  \Phi \, .
\ea

We need to get rid of the last term above containing second derivatives of
$n$. To do this, we take derivatives of the identity $G(n,n)=1$ in GNC,
which leads to
\ba \label{nab}
\ga^{ab} G_{MN} n^{M} n^{N}_{, ab} \, = \,
 - \frac{1}{2} n^{M} n^{N}\, ( G^{PQ} -n^{P} n^{Q} ) G_{MN, PQ} -
K^{ab}K_{ab} \, .
\ea

Using the above expression in Eq. (\ref{deltaKab2}), we have
\ba \label{deltaKab3}
\ga^{ab}\delta K_{ab} \, = \, - \nabla^{2} \Phi +K^{ab}K_{ab} -
R_{MN} n^{M} n^{N} \, ,
\ea
where $R_{MN}$ is the Ricci tensor of the background geometry.

Thus, our final expression for $\delta (LHS)$ in Eq. (\ref{LHS1}) is
\ba \label{LHS2}
\delta (LHS) \, = \, \sqrt{-|\ga| } \left [ \,  - \nabla^{2} \Phi +
( K^{2} -  K^{ab} K_{ab}  - R_{MN} n^{M} n^{N}  ) \, \Phi  \, \right ]
\ea
in agreement with  \cite{Guven:1993ew}.

For the RHS of Eq. (\ref{eom1}) we have
\ba \label{RHS1}
 \delta(RHS) \, = \, \left [ C_{(4)}'' n^{r} (n^{r} - \dot R\,  n^{0} ) +
C_{(4)}' (n^{r} \dot {n^{0}} - n^{0} \dot{ n^{r} } ) \right] \,  \Phi+
C_{(4)}' (\delta n^{r} - \dot R\, \delta n^{0} )\, \Phi \, .
 \ea

Using the formula (\ref{deltan}) for $\delta n$, one can show that
\ba \label{RHSlast}
\delta n^{r} - \dot R \, \delta n^{0} \, &=& \, \frac{(G_{00} + G_{rr} \dot R^{2} )\,  n^{r} }{2 G_{00}\, g_{rr}\,
(n^{0} \dot R - n^{r} ) } \left(\,  G_{00}' \, ( n^{0})^{2} + G_{rr}' \, ( n^{r})^{2}  \, \right)  \Phi\nonumber\\
&=& \, \frac{n^{r} }{\dot R} ( \dot{n^{r}} - \dot R \, \dot{n^{0}} ) \, ,
\ea
where to get the final line the relations $G(n,n)=1$ and  $G(n, e_{a})=0$ were used.

Combining Eqs. (\ref{RHSlast}) and (\ref{RHS1}) we obtain
\ba \label{RHS2}
\delta(RHS) \, = \,
( \frac{n^{r} }{\dot R}  -n^{0})\,  (C_{(4)}' n^{r} )_{,t} \Phi \, .
\ea

Combining Eqs. (\ref{RHS2}) and (\ref{LHS2}), the desired linear
equation of motion from Eq. (\ref{eom1}) is
\ba
\nabla^{2} \Phi + \left[ K_{ab} K^{ab} +R_{M N} n^{M} n^{N} - K^{2}
-\frac{1}{\sqrt{-|\ga| }}( n^{0} - \frac{n^{r} }{\dot R} )\,  (C_{(4)}' n^{r} )_{,t} \right] \Phi \, = \, 0 \, .
\ea

\section{Approximations for $I(r)$ and $K(r)$}
The following approximation formulae for the KS background are useful:
\ba
\label{asym}
K(r \rightarrow 0) \, &\rightarrow& \, (2/3)^{1/3}+{\cal{O}}(r^2) \, , \nonumber\\
K(r \rightarrow \infty) \, &\rightarrow& \, 2^{1/3}\, e^{-r/3} \, ,\nonumber\\
I(r \rightarrow 0) \, &\rightarrow& \, 0.72 +{\cal{O}}(r^2)\, , \nonumber \\
I(r \rightarrow \infty) \, &\rightarrow& \, 3\, .\, 2^{-1/3}
\left(r-\frac{1}{4} \right) e^{-4r/3}  \, .
\ea

\end{document}